# Upgrade of the Data Acquisition and Control System of Microwave Reflectometry on the Experimental Advanced Superconducting Tokamak

Fei Wen, Haoming Xiang, Tao Zhang, Yuming Wang, Xiang Han, Hao Qu, Fubin Zhong, Kaixuan Ye, Mingfu Wu, Gongshun Li, Shoubiao Zhang, Xiang Gao

*Abstract*—The reflectometry on Experimental Advanced Superconducting Tokamak (EAST) is undergoing an upgrade for more comprehensive measurement of plasma density profile and fluctuation. The Data Acquisition and Control System (DACS) has been redeveloped to satisfy the requirements of the upgraded reflectometry. The profile reflectometry works in 30-110 GHz (X-mode) and 40-90GHz (O-mode), when the fluctuation reflectometry operates at 20 fixed frequency points in 50-110GHz (X-mode) and 20-60GHz (O-mode). The PXIe-based DACS includes two 8-channel 14-bit 250MSPS digitizers and ten 8-channel 12-bit 60MSPS digitizers. A self-developed 5-channel 250MSPS arbiter waveform generator (AWG) is used to control voltage control oscillators for frequency sweeping. A trigger and clock manger and a timing module receive the trigger and clock signal from central controller and synchronize all the digitizers and the AWG. The total data rate from digitizers is 2515 MB/S. The Data from digitizers is streamed to a disk array (RAID 0) with data throughput capacity of 3000 MB/S. Meanwhile, selected data is transported to a FPGA based real-time data processing module, which utilize a pre-trained neural network to calculate the plasma density profile. Now the new reflectometry is being installed on EAST, and its performance will be tested in experimental campaign after 2018.

## I. INTRODUCTION

MICROWAVE reflectometry is an important diagnostic tool for plasma density measurement[1]–[3] which is widely used on fusion devices, including JET[4], [5], Tore-Supra[6], DIII-D[7], [8], ASDEX Upgrad[9], [10], and so on. On the Experimental Advanced Superconducting Tokamak (EAST) development of reflectometry system was started in 2012. The swept frequency reflectometries that operate in Q, V and W band and the fixed frequency reflectometry working in V band were gradually installed on the EAST between 2012 and 2016[11]–[14]. By 2017 the reflectometry on EAST could provide plasma density profile and fluctuation measurement sinuously in the pedestal[15], [16].

To extend the detection range from pedestal to core and provide more comprehensive measurement, the reflectometry on the EAST is undergoing an upgrade. The upgraded reflectometry system consists of 9 reflectometries as shown in Table I. There are four fixed frequency reflectometries for plasma density fluctuation measurement, which works at 8 fixed point in 20-60 GHz (O mode) and 12 fixed point in 50-75 GHz(X-mode). Meanwhile, 5 swept frequency reflectometry for plasma density profile operate in 23-110GHz (X mode) and 40-90 GHz (O mode). The reflectometries can work independently, but they might share transmission lines, waveguides or antennas because of the lack of diagnostic port and limited space.

TABLE I
OPERATING BAND AND MODE OF UPGRADED REFLECTOMETRY SYTEM

| Operating Band | Mode | Frequency | Target |
|---|---|---|---|
| K and Ka band (20-40 GHz) | O mode | Fixed | Fluctuation |
| U band (40-60 GHz) | O mode | Fixed | Fluctuation |
| V band (50-75 GHz) | X mode | Fixed | Fluctuation |
| W band (75-110 GHz) | X mode | Fixed | Fluctuation |
| Q band (32-56 GHz) | X mode | Swept | Profile |
| V band (48-76 GHz) | X mode | Swept | Profile |
| W band (72-110 GHz) | X mode | Swept | Profile |
| U band (40-60 GHz) | O mode | Swept | Profile |
| E band (60-90 GHz) | O mode | Swept | Profile |

In the following chapters, an upgraded Data Acquisition and Control System (DACS) for the reflectometry system will be introduced. The new DACS is designed to meet the

This work was supported by the National Key R&D Program of China (No. 2014GB106003), and the National Natural Science Foundation of China (No. 11675211, No.11605235, No. 11505221).

Fei Wen, Tao Zhang, Yuming Wang, Xiang Han, Shoubiao Zhang, Xiang Gao are with Institute of Plasma Physics, Chinese Academy of Sciences, PO Box 1126, Hefei, Anhui 230031, People's Republic of China (e-mail:wenfei@ipp.ac.cn).

Haoming Xiang, Hao Qu, Fubin Zhong, Kaixuan Ye, Mingfu Wu are with Institute of Plasma Physics, Chinese Academy of Sciences, PO Box 1126, Hefei, Anhui 230031, People's Republic of China. They also are with University of Science and Technology of China, Hefei 230026, P. R. China.

Gongshun Li is with Advanced Energy Research Center, Shenzhen University, Shenzhen 518060, People's Republic of China. He is also with Key Laboratory of Optoelectronic Devices and Systems of Ministry of Education and Guangdong Province, College of Optoelectronic Engineering, Shenzhen University, Shenzhen 518060.

requirements not only from the current upgrade but also from the future improvement of reflectometry system on EAST.

## II. REQUIREMENTS OF REFLECTOMETRY SYSTEM

A typical architecture of microwave reflectometry on the EAST is shown in Fig. 1. In the transmitter of reflectometry, a voltage control oscillator (VCO) or phase locked oscillator (PLO) with output multiplied is used as microwave source. The microwave signal is amplified and lunched to plasma after part of its power is coupled to the receiver as reference beam. The microwave is reflected at cut-off layer in plasma and is collected by the receiving antenna. The reflected signal is mixed with output signal of local oscillator and then is demodulated by in-phase and quadrature detector (I/Q detector). The in-phase and quadrature signal (I/Q signal) obtained from the I/O detector is acquired by the DACS.

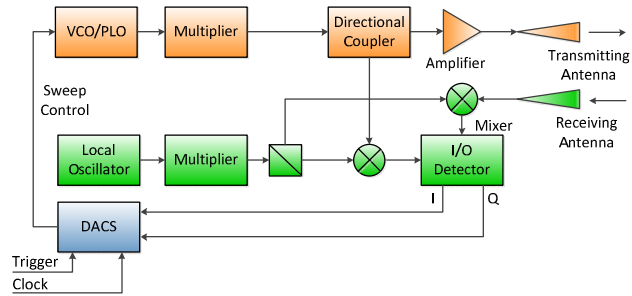

Fig. 1. Typical architecture of reflectometry

The sampling frequency of the DACS system is determined by the I/Q signal. For plasma density fluctuation measurement, the I/Q signal's frequency is only related to the frequency of plasma fluctuation. On the EAST the fluctuation frequency that we are interested in is no more than 1 MHz. As a result, the sampling rate for fluctuation measuring should be no less than 2 MSPS.

For plasma density profile measurement with the swept frequency probing, the I/Q signal's frequency is determined by the beat frequency ($F_{beat}$), which is a function of sweeping time and propagation time($\tau$) of the microwave into the plasma,

$$F_{beat} = \tau \cdot \frac{dF}{dt}$$

In current experimental setup, the sweeping time is 40 μS with 10 μS dead time. In the reflectometry a delay line is used on reference beam to reduce the impact of propagation time on beat frequency. Currently the beat frequency is no more than 10MHz, but in the future the sweeping time might be reduced to 10uS to provide a better temporal resolution. The DACS needs fast enough sampling rate to accommodate the beat frequency.

To monitor transient events in the tokamak, including edge local modes (ELMs), L mode to H mode transient, I phase and so on, the DACS should continuously record all the signal during a shoot. The EAST aims for long plasma pules of up to 1000 seconds. On July 3, 2017, EAST became the first tokamak to successfully sustain H-Mode plasma for over 100 seconds. The DACS should have enough continues data streaming capability and massive data storage capability.

Under the swept frequency mode, the reflectometry need a voltage control signal (0-20V) to modulate the VCO. The control signal should be a carefully calculated curve to calibrate the nonlinear of the VCO. To avoid bringing more noise, an arbitrary waveform generator(AWG) providing output signal with high-precision time and amplitude is necessary.

The reflectometry works with lots of other diagnostics tools on the EAST. To ensure data synchronization, the DACS need to work under the control of trigger signal and clock signal from the central control. Due to the delay caused by the long transmission line, the trigger need to be careful adjusted to provide perfect synchronization.

## III. DETAILED DESIGN OF THE DACS

The DACS is based on PXIe bus. As shown in Fig.2, the digitizers, timing module, data processor, RAID controller with disk array are all connected to controller by PXIe bus. The AWG is controlled by the controller via Universal Serial Bus (USB). A trigger and clock manager (TCM) provides the synchronization for the DACS. The DACS interact with the other controllers and upload data to data warehouse by gigabit ethernet. Most devices in the DACS are commercial off-the-shelf (COTS) components except the AWG and trigger & clock manager.

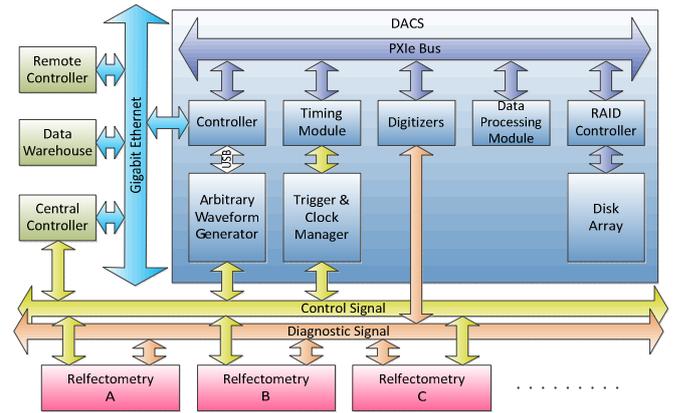

Fig. 2. Architecture of the DACS

### A. Digitizer

For measuring the density fluctuation with poloidal and radial correlation property, the reflectometry system operates at 20 fixed frequency points in 20-60 GHz (O mode), and 50-110 GHz (X mode), and uses 2 receiving antennas at different poloidal position for each frequency point. So, there are 40 receivers with I/Q demodulators outputting 80 I/Q signals. The DACS applies ten 8-channel 12-bit 60 MSPS digitizers (PXIe-5105) to acquire the I/Q signals. The PXIe-5105 provides a programmable input vertical range from 0.05V to 6V with input impedance of 50Ω or 1MΩ. The Signal to Noise and Distortion (SINAD) is 61 dB when input range is set to 1V. The sample rate can be set between 4 MSPS and 60 MSPS when the external sample clock is used. The digitizer PXIe-5105 can provide ideal performance to acquire the signal with frequency of 1MHz from the fluctuation reflectometry.

For plasma density profile measurement, 5 swept frequency reflectometries working in 32-100 GHz (X mode) and 40-90

GHz (O mode) have 5 receivers and 10 I/Q signals to be acquired. Two 8-channel 14-bit 250MSPS (PXIe-5171R) are implemented in the DACS. The PXIe-5171R can only support 50Ω input impedance and input voltage range of 0.2-5V. The Analog-to-Digital Converter (ADC) work at a fixed rate of 250MSPS, but decimation factor can be set to get lower sample rate. The beat frequency of the profile reflectometry is lower than 10MHz, so decimation factor is set as 4 and the sample rate is 62.5 MSPS. The Effective Number of Bits (ENOB) is 11 with anti-alias filter enabled and input frequency lower than 30MHz. The digitizer also provides a Field-programmable gate array (FPGA Xilinx Kintex-7 XC7K410T) on board to customize the functions of digitizer. In addition to 10 acquisition channels for output of reflectometry, there are still 6 spare channels. 5 of them are used to acquire and monitor the sweep control signal from the AWG.

### B. Arbitrary Waveform Generator

The profile reflectometries need 5 VCO control signals. If we choose commercial signal generators, 3 or more instruments should be installed in the control cubical which is too crowded to accommodate them. So, a self-developed AWG is applied in the DACS.

The AWG utilizes a Cyclone IV FPGA as controller. An USB interface chip provide the communication between the FPGA and the DACS controller. The order and waveform data are transported via USB and stored in the internal memory of the FPGA. Three dual 14-bit 250MSPS Digital-to-Analog Converters (DAC AD9746) are chosen to generate the waveform. The output of DAC (-1V to 1V) is low-pass flited and amplified by three cascaded amplifiers to attain the output range of 0-20V as sweep control signal. Meanwhile, a reduced copy of sweep control signal (0-1V) is sent to the digitizer for monitoring.

To reduce the non-linear noise, a look-up table storing the calibration curve is built in the FPGA. The non-linear properties of DAC and amplifiers are considered in the calibration curve.

### C. Synchronization

To provide perfect synchronization, the clock and trigger signal from central controller are conditioned by a self-developed TCM. The ADCs and DACs in the DACS all require high-quality clock (< 1 ps jitter rms). The input clock is connected to a jitter cleaner (CDCE62002) which can achieve good jitter performance under 0.5 ps rms. The cleaned clock is directly connected to the DAC in the AWG and the timing module in the DACS.

The trigger from central controller arrives different diagnostic tools at different times that will cause the data from different source can not align with each other. To align the trigger signals, delay lines are used to adjust the trigger delay between 20 nS to 500 nS. The delayed triggers are sent to the AWG and the timing module.

The sweep frequency monitoring signal from the AWG is directly connected to digitizer, while the output signal of the VCO need propagation time to pass through the microwave devices, waveguide and cables. As a result, there is misalignment in the data between sweep frequency monitoring signals and I/Q signals. The time gap between them is about 200ns, which will cause the frequency error of 100-200MHz in data processing. To resolve the problem, the sweep frequency monitoring signal is also delayed by the delay line to align with the I/Q signals before it is sent to digitizer.

The 12 digitizers are not directly synchronized by the TCM, but by the timing module connected to PXIe Bus. The timing module provides PXI_CLK10 signal from the oscillator on board to drive the digitizers as clock source. After the timing module receive the clock from the TCM, the PXI_CLK10 is automatically phase locked to the input clock. An independent buffer with a source impedance matched to the backplane is applied for the PXI_CLK10 signal, and skew of slots is less than 1ns. The PXIe bus provides PXI Differential Star Triggers (PXIe-DSTAR) whose slot to slot skew is only 150 ps. The PXIe-DSTAR is applied to distribute the trigger signal from the TCM to digitizers.

### D. Data Streaming

The 12 digitizer generates massive of data. The total data rate ($D_{total}$) can be calculated by the equation,

$$D_{total} = w(nS_H + mS_L)$$

Where n is the number of high-speed channels (15); m is number of low-speed channels (80); $S_H$ is sample rate of high-speed channel (62.5 MSPS); $S_L$ is sample rat of low-speed channel (4 MSPS); $w$ is the width of data in byte (2). The total date rate of the DACS is 2515 MB/S. With this data rate, the DACS needs to work continuously for hundreds of seconds.

The high data rate is the main reason we choose PXIe bus. The PXIe chassis (PXIe-1085) provide 24 GB/S total data bandwidth and 8 GB/S data bandwidth for one slot. The data width is only theoretical value, not considering the time spent on data packaging and transport protocol, but the maximum effective bandwidth, about 85% of theoretical one, already well meet the transmission requirements.

The transmitted data should be promptly stored, and the storage speed must exceed the total data rate of 2515 MB/S. A disk array of RAID0 (Redundant Array of Independent Disks) is implemented to store the high rate data. The disk array consists of 24 250 GB solid-state drives (SSD). The coming data is spited into 24 parts and stored sinuously. The test has shown that the maximum data write speed is more than 3000 MB/S. The disk array provide data storage capability of 5.6 TB that means the DACS can continuously acquire data of over 2300 seconds.

### E. Data Processing

To get the plasma density profile in real time, which is important for the plasma position control, a Data Processing Module (DPM) is added to the DACS. The DPM support peer to peer streaming, that means the data can directly transfers between the digitizers and the DPM without sending data through the controller. This technology improves real-time performance of data processing and diminishes the impact on data acquisition. Considering the data processing capability, only small partial data (extract 50 uS data every 1 mS) will be transported the DPM. The DPM utilizes a neural network to calculate the density profile which is much faster than classical inversion calculation. The neural network is pretrained on the computer with massive historical data to obtain various parameters. After the parameters of neural network are all confirmed, it is implemented in the FPGA of the DPM. The FPGA is very suitable for implementing this kind of parallel computing. Although the accuracy stilled need to be optimized, the time for obtaining density profile has reduced to 1 mS.

## IV. Conclusion

For the upgrade of reflectometry system on the EAST, the DACS's capability of various aspect is improved. The number high-speed data acquisition channels increase from 6 to 16, when the number of low-speed ones increase from 32 to 80. The AWG's sample rate is improved to 250MSPS with 5 outputs. The data streaming capability reaches 3000 MB/S. The data storage capability is 5.6 TB. The real time data processing is tried and can offer density profile in 1 mS. All the improvements make the DACS to be qualified for the reflectometry data acquisition work on the EAST.

Part of reflectometry system is still being installed, so we haven't got any actual experimental data. The performance of the upgraded reflectometry system will be seen in experiments after 2018.